\documentclass[journal=jacsat,manuscript=article]{achemso}

\usepackage[version=3]{mhchem} 
\usepackage[usenames,dvipsnames]{xcolor}
\usepackage{float} 

\author{Kamal Choudhary}
 \email{kamal.choudhary@nist.gov}
 \affiliation{%
 Material Measurement Laboratory, National Institute of Standards and Technology,
Gaithersburg, MD 20899, USA 
}
\title{AtomGPT: Atomistic Generative Pre-trained Transformer for Forward and Inverse Materials Design }

\abbreviations{IR,NMR,UV}

\begin{document}

\begin{tocentry}




\begin{center}
\includegraphics[width=8cm]{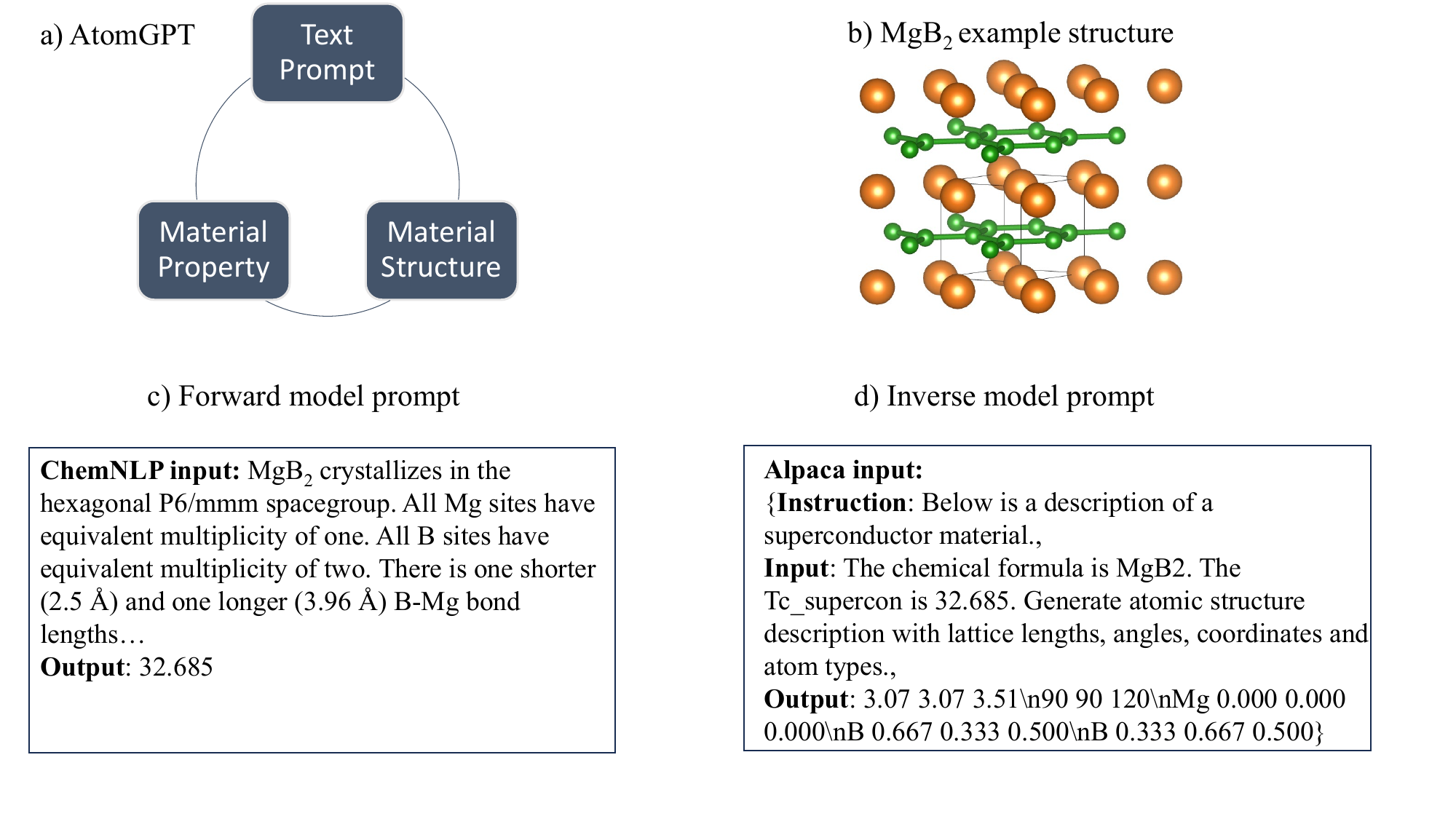}
\label{TOC}
\end{center}
\end{tocentry}

\begin{abstract}
Large language models (LLMs) such as generative pretrained transformers (GPTs) have shown potential for various commercial applications, but their applicability for materials design remains underexplored. In this article, AtomGPT is introduced as a model specifically developed for materials design based on transformer architectures, demonstrating capabilities for both atomistic property prediction and structure generation. This study shows that a combination of chemical and structural text descriptions can efficiently predict material properties with accuracy comparable to graph neural network models, including formation energies, electronic bandgaps from two different methods, and superconducting transition temperatures. Furthermore, AtomGPT can generate atomic structures for tasks such as designing new superconductors, with the predictions validated through density functional theory calculations. This work paves the way for leveraging LLMs in forward and inverse materials design, offering an efficient approach to the discovery and optimization of materials.
\end{abstract}


In recent years, machine learning (ML) has emerged as a powerful tool in materials science, offering the potential to accelerate materials discovery by learning from existing data and making predictions for new materials \cite{choudhary2022recent,vasudevan2019materials,schmidt2019recent}. Such ML models can be used for forward design tasks such as given an atomic structure predict its properties, or inverse design tasks such as given properties predict its atomic structures. Among various ML approaches such as descriptor-based, graph-based and text-based methods, large language models (LLMs) have shown promises to the exploration of vast materials design space beyond the limitations of traditional methods\cite{jablonka202314,choudhary2023artificial,goodfellow2016deep}.

The Generative Pre-trained Transformer (GPT) is a type of large language model (LLM) originally developed for natural language processing and has demonstrated remarkable success in generating coherent and contextually relevant text \cite{vaswani2017attention,tunstall2022natural,rothman2021transformers}. Models such as ChatGPT \cite{wu2023brief} have been used for code generation/debugging, literature review and numerous other tasks. However, if we attempt to perform forward/inverse materials design tasks, the outcomes can be quite poor\cite{pimentel2023challenging,jablonka2024leveraging,polak2024extracting}. Nevertheless, inspired by its simplicity of usage and the massive success of ChatGPT, an alternate model AtomGPT (Atomistic Generative Pre-trained Transformer) is introduced tailored for forward and inverse materials design.

There have been several previous research works for forward prediction material properties such as descriptor-based methods including models in MatMiner \cite{ward2018matminer}, Classical Force-field Inspired Descriptors (CFID) \cite{choudhary2018machine}; graphs-based methods such as Crystal graph convolutional neural networks (CGCNN) \cite{xie2018crystal}, MatErials Graph Network (MEGNet) \cite{chen2019graph}, Atomistic Line Graph Neural Network (ALIGNN) \cite{choudhary2021atomistic}, \textcolor{black}{CoGN\cite{reiser2021graph}, M3GNet \cite{chen2022universal}, Matformer \cite{yan2022periodic}, ComFormer \cite{yan2024complete}}; and language-based methods such as LLM-Prop \cite{rubungo2023llm} etc. Similarly, some of the earlier inverse design methods are crystal diffusional variational autoencoder (CDVAE) \cite{xie2021crystal}, Fourier-transformed Crystal Properties (FTCP) \cite{ren2020inverse}, G-SchNet \cite{gebauer2019symmetry}, MatBERT \cite{korolev2023accurate}, \textcolor{black}{CrystaLLM} \cite{antunes2023crystal}, Crystal-LLM \cite{gruver2024fine}, and xyztransformer \cite{flam2023language}.

AtomGPT is a deep learning model \cite{goodfellow2016deep} that leverages a few transformer architectures such as GPT2 \cite{radford2019language} and quantized Mistral-AI \cite{jiang2023mistral} to learn the complex relationships between atomic structures and material properties from dataset such as JARVIS-DFT \cite{choudhary2020joint,wines2023recent,choudhary2018computational}. JARVIS-DFT includes a diverse range of structural, electronic, and thermal properties. By learning from this dataset, AtomGPT acquires a rich representation of the materials space, enabling it to predict properties of unseen materials and generate new material structures with desired properties.

One of the key features of AtomGPT is its ability to perform both forward and inverse materials design. In forward design, the model predicts the properties of a given material structure, allowing researchers to screen and evaluate potential materials efficiently. In inverse design, AtomGPT generates material structures that are likely to exhibit specific target properties, guiding the synthesis of materials with desired characteristics.  
\textcolor{black}{The forward model is intended for pre-screening purposes, where it efficiently predicts properties of known structures. This allows us to quickly filter through a vast number of potential candidates and identify those with desirable properties. This step is crucial in narrowing down the search space and ensuring that only the most promising candidates are considered for further analysis. On the other hand, the inverse models are designed for structure generation based on desired properties. This complementary approach to funnel-type screening allows us to directly generate new structures that meet specific criteria. By focusing on this goal, the inverse models can explore the structural space more effectively and generate novel candidates that may not have been considered otherwise. By maintaining separate models for these tasks, we can leverage the strengths of each approach without compromising their effectiveness. The forward model excels in rapid property prediction, while the inverse models specialize in creative structure generation. This dual capability makes AtomGPT a versatile tool for materials scientists, enabling them to navigate the materials space more effectively and accelerate the discovery of novel materials.}

Additionally, as AtomGPT is using transformers and other cutting-edge LLM library models, it can integrated with newer models as they come into existence. Finally, AtomGPT is closely integrated with NIST-JARVIS infrastructure\cite{wines2023recent,choudhary2020joint}, which has several existing databases, streamlined workflows, benchmarks and models to evaluate them using a detailed materials domain knowledge base. The AtomGPT model training scripts will be made available at \url{https://github.com/usnistgov/atomgpt}.

AtomGPT's application potential is vast, spanning various domains of materials science. In this work, the focus is going to be on designing Bardeen–Cooper–Schrieffer (BCS)-superconductors \cite{cooper2010bcs,giustino2017electron,choudhary2022designing} with relevant material properties for forward and inverse methods. Finding superconductors with high transition temperatures ($T_C$) at ambient as well as extreme conditions has been an active area of research for decades\cite{cooper2010bcs}. Predicting such quantities using experiments and computational methods such as density functional theory \cite{giustino2017electron} can be time-consuming and artificial intelligence-based methods \cite{choudhary2022designing,schmidt2023machine} can be helpful to accelerate the design process. In addition to $T_C$, some of the other properties of interest for superconductor design are formation energy (should be negative) and electronic bandgap (should be close to zero to follow BCS formalism), which are also considered in this work.






\begin{figure}[hbt!]
\centering
\includegraphics[width=\linewidth]{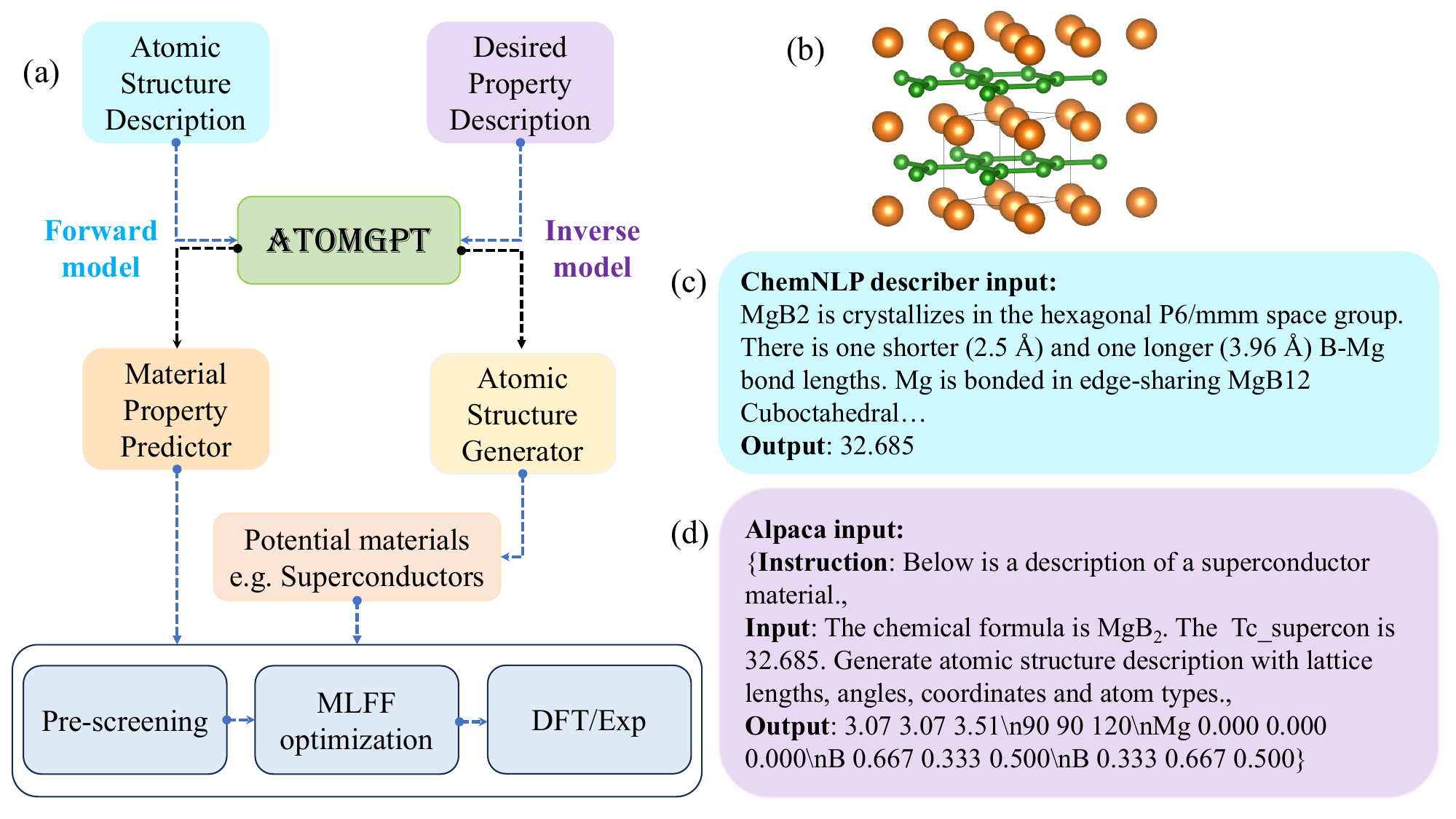}
\caption{Schematic overview of the AtomGPT workflow. AtomGPT can be used for both forward model (atomic structure to property) predictions as well as inverse design (property to atomic structure generation) using LLMs. a) integrated text to material property predictions, text input to atomic structure generation, pre-screening, unified machine learning force-field (MLFF) optimization and density functional theory (DFT) calculations/experiment (Exp) based validation processes b) an example crystal structure for BCS-superconductor MgB$_2$ (JVASP-1151), c) text description of atomic structure of MgB$_2$ including explicit atomic structure as well as chemical information using ChemNLP, d) text prompt to explicit atomic structure generation example using Alpaca format.}
\label{fig:zsl}
\end{figure}

An overview of the AtomGPT package is shown in Fig. 1a. AtomGPT can be used to predict the property of a material given its information or generate its structure given its property information in terms of text. To illustrate these processes, an example for well-known BCS superconductor MgB$_2$ 
(atomic structure shown in Fig. 1b) with an experimental $T_C$ of 39 K \cite{tomsic2007overview}, and DFT prediction of 32.7 K \cite{choudhary2022designing} is shown in Fig. 1c and Fig. 1d. In order to represent the chemical and structural information of materials ChemNLP describer\cite{choudhary2023chemnlp} or Robocrystallographer \cite{ganose2019robocrystallographer} can be used. Such a description includes instructions such as chemical formula, space group, bond-distance, bond-angles and explicit fractional coordinates of a material as shown in Fig. 1c. 

First, let us discuss the method involved in forward property prediction models with a GPT2 architecture available in the transformers library \cite{wolf2019huggingface}. AtomGPT allows easy integration of any other related huggingface library models available in the transformers library to be easily used in place of the GPT2 model \cite{radford2019language}. GPT2 was developed by OpenAI and its successors GPT3.5 and GPT4 are now commercial and have revolutionised the field of artificial intelligence. GPT-2 was primarily trained on CommonCrawl dataset and utilizes a stacked transformer decoder architecture, which consists of multiple layers of transformer blocks  \cite{radford2019language}. Each transformer block contains two main components: a multi-head self-attention mechanism and a position-wise feed-forward network. The input to the model is a sequence of tokens, which are first converted into embeddings and then passed through the transformer blocks. Scaled dot-product attention used in a transformer model can be written as:
\begin{equation}
   \text{Attention}(Q, K, V) = \text{softmax}\left(\frac{QK^T}{\sqrt{d_k}}\right)V
\end{equation}
where \(Q\), \(K\), and \(V\) represent the query, key, and value matrices, respectively. Here, \(d_k\) is the dimensionality of the key vectors. The multi-head attention is obtained by concatenating multiple such attention heads. The multi-head self-attention mechanism allows the model to focus on different parts of the input sequence when computing the output for a particular token.

As the LLM models such as GPT2 were primarily designed for text to text models and not explicitly for regression applications, the language model head is modified with a linear neural network model to predict one or several properties as output(s). A pre-trained GPT-2 is loaded as a causal language model (CausalLM), which means it's designed for tasks where the prediction of each word token depends only on the previous tokens in the sequence (e.g., text generation). The new network with a modified language model head is a sequential model consisting of two linear layers. The first linear layer maps the hidden states of the GPT-2 model to a lower-dimensional latent space dimension (here 1024). The second linear layer maps this latent representation to a single output value for a regression task such as formation energy or superconductor $T_C$. There are various versions of GPT-2 model, but in this work, GPT2-small is used with 124 million parameters for forward design tasks. The models are trained using V100 GPUs with a batch size of 10, L1 loss function, and a learning rate of 10$^{-3}$ for 200 epochs. The hyperparameters such as batch size, learning rate, latent space dimension were chosen by varying them for the smallest superconductor database, and used for other tasks with larger datasize such as formation energy and bandgaps.

The dataset used for this work is taken from the JARVIS-DFT and includes formation energies (55713), OptB88vdW bandgaps (55713), Tran Blaha modified Beck-Johnson potential based bandgaps (18167) and superconducting transition temperatures (1058) at ambient pressures \cite{choudhary2020joint,wines2023recent,choudhary2018computational}. The dataset provides atomic structure information such as atomic coordinates, constituent element types and lattice parameters along with DFT calculated properties mentioned above. The dataset is used for 80:10:10 splits for training, validation and testing splits. 

As LLM models require text as inputs, we can provide a description of an atomic structure using ChemNLP \cite{choudhary2023chemnlp} which is then fed to the model discussed above. \textcolor{black}{The input of the LLM-based crystal property prediction is carried out with ChemNLP describer \cite{choudhary2023chemnlp} which is based on well-defined combinations of structure and chemical features based on JARVIS-Tools. It generates inputs in a deterministic manner, ensuring that the same crystal structure consistently results in the same input description.} 


\begin{figure}[H]
\centering
\includegraphics[width=\linewidth]{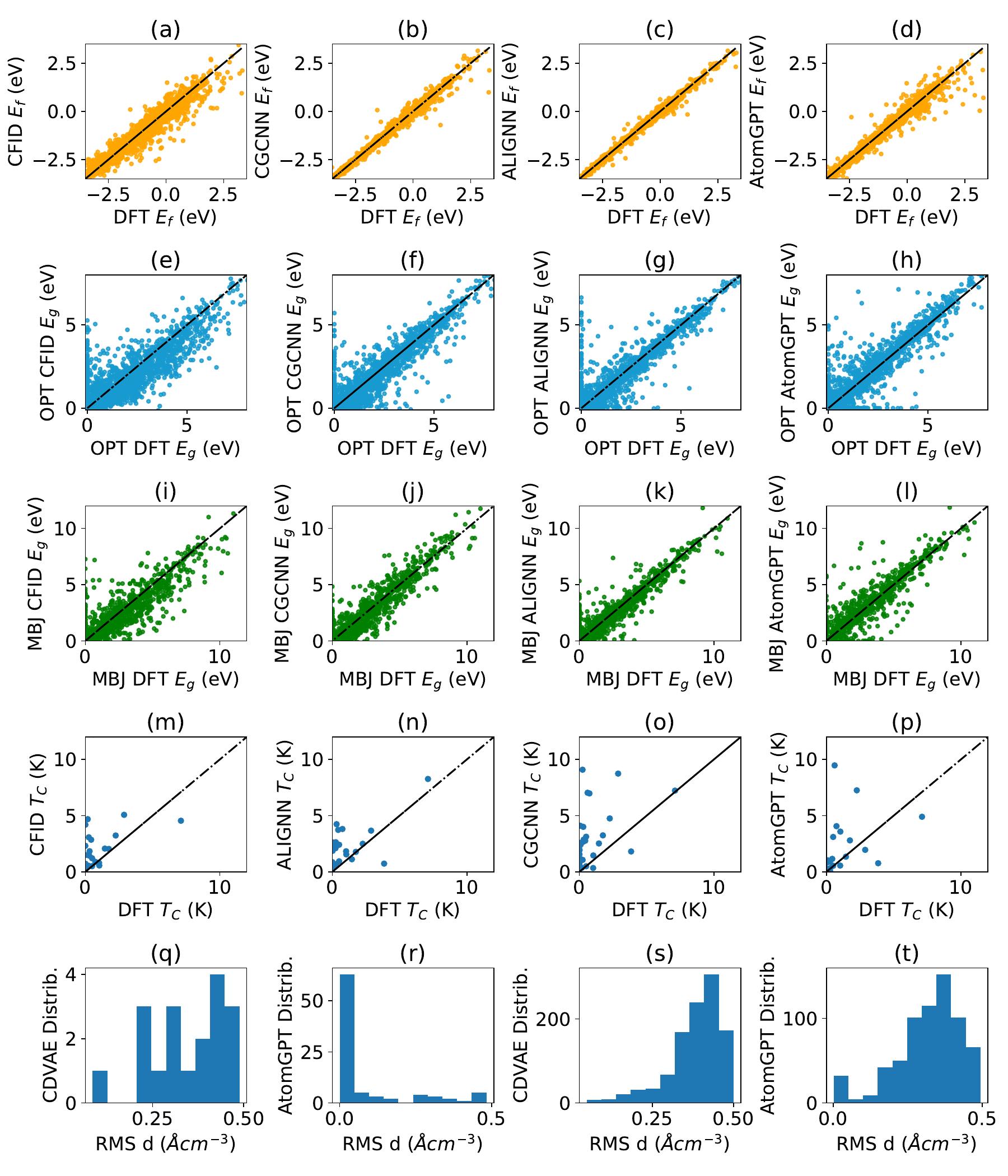}
\caption{ Performance of various models for forward (a-p) and inverse design (q-t) tasks on held out test set. Forward model performance for formation energy ($E_f$), OptB88vdW (OPT) bandgap ($E_g$, TBmBJ (MBJ) bandgap, and superconducting transition temperature ($T_C$) are shown using CFID, CGCNN, ALIGNN and AtomGPT models. \textcolor{black}{Similarly, inverse model performance for superconductor dataset (q-r) and Carbon energy dataset (s-t) are shown using CDVAE and AtomGPT models.} }
\label{fig:scatter}
\end{figure}

After training such models, the performances are shown in Table 1 and Fig. \textcolor{black}{2a-2p} for the forward design models of formation energy per atom, OptB88vdW bandgaps, TBmBJ bandgaps and superconducting transition temperatures using identical train-test splits (available in the JARVIS-Leaderboard benchmarking platform\cite{choudhary2024jarvis}) for CFID, CGCNN and ALIGNN methods. \textcolor{black}{For formation energy model the mean absolute error (MAE) values for CFID, CGCNN, ALIGNN and AtomGPT are 0.142 eV/atom, 0.063 eV/atom, 0.033 eV/atom and 0.072 eV/atom respectively. The mean absolute deviation (MAD), which can serve as a model baseline, is computed by using the mean of the target values in the training dataset and using it as predictions for all the materials in the test dataset. Usually, MAD:MAE of greater than 1 is desirable and higher values suggest better models. \textcolor{black}{We note that while a MAD:MAE of greater than 1 is desirable, a higher ratio could also indicate a significant impact from outliers and data skewness}. The MAD:MAE for the above models are 6.05, 13.74, 25.94 and 11.54 respectively. For OptB88vdW semi-local functional based electronic bandgap, the MAE values for CFID, CGCNN, ALIGNN and AtomGPT are 0.299 eV, 0.199 eV, 0.142 eV and 0.139 eV respectively. The MAD:MAE values are 3.31, 5.00, 6.97 and 7.11. For TBmBJ meta-GGA based electronic bandgap, the MAE values for CFID, CGCNN, ALIGNN and AtomGPT are 0.531 eV, 0.407 eV, 0.310 eV and 0.319 eV with MAD:MAE values of 3.35, 4.37, 5.73, 5.58 respectively. For electron-phonon coupling calculation based superconductor transition temperatures, the MAE values for CFID, CGCNN, ALIGNN and AtomGPT are 1.99 K, 2.06 K, 2.03 K and 1.54 K with MAD:MAE values of 0.890, 0.687, 0.880 and 1.160 respectively.} Based on the above results, AtomGPT outperforms ALIGNN for OptB88vdW bandgap and $T_C$ while for TBmBJ bandgaps and formation energies, ALIGNN still dominates other models. \textcolor{black}{Lower MAE values in Table 1 and data along x=y line in scatter plots (Fig. 2a-2p) represent better performance.} The superior performance of ALIGNN for formation energy can be attributed to some of the sophisticated information such as bond-angles, directional graphs and atom feature information \cite{choudhary2021atomistic} which are not explicitly included in the AtomGPT model, Nevertheless, AtomGPT performance for formation energy is closer to CGCNN which is also a GNN model with sophisticated graphs for atomistic systems. Note, it took dedicated efforts to develop such domain specific models such as ALIGNN, while finetuning GPT2 model is very straightforward. For some cases such as TBmBJ, the performances of ALIGNN and AtomGPT are comparable which is promising. 

\textcolor{black}{It is important to highlight the distinct design philosophies and underlying mechanisms of ALIGNN and AtomGPT. ALIGNN is a sophisticated graph neural network (GNN) that incorporates intricate features such as bond angles, directional graphs, and specific atom features, enabling it to excel in predicting properties such as the formation energy. The development of ALIGNN required extensive efforts to finely tune these domain-specific features, resulting in its superior performance for formation energy predictions. On the other hand, AtomGPT is based on a transformer architecture, specifically designed to leverage the strengths of large language models (LLMs). While fine-tuning AtomGPT is indeed straightforward compared to the extensive development process of ALIGNN, AtomGPT benefits from the ability to learn complex relationships between atomic structures and material properties through its deep learning capabilities. AtomGPT's transformer-based architecture allows it to generalize well across diverse datasets. This could be particularly beneficial for predicting $T_C$, where complex electronic interactions play a significant role. ALIGNN's explicit incorporation of sophisticated structural features gives it an edge in tasks like formation energy prediction, where detailed geometric and chemical interactions are critical. In contrast, AtomGPT relies on textual and structural descriptions, which, while powerful, may not capture all nuances as effectively for certain properties.}\textcolor{black}{We note that using LLM/GPT for materials datasets is a nascent area of research and analyzing the strength/weakness of such models for materials design need further explorations.}


Next, the inverse design task is demonstrated by developing a separate model. While the forward prediction model uses GPT2 with a modified large model head, the inverse design model is based on supervised fine-tuning with Mistral-AI 7 billion parameter model \cite{jiang2023mistral} using Low-Rank Adaptation (LoRA) for parameter-efficient fine-tuning (PEFT) \cite{hu2021lora}. Mistral is a powerful model with 7.3 billion parameters and has been shown to outperform Large Language Model Meta AI (LLaMA) 2 13B \cite{touvron2023llama2}, LLaMA 1 34B \cite{touvron2023llama}, and ChatGPT3.5 \cite{wu2023brief} on several publicly available benchmarks. Some of the key features of Mistral are: sliding window attention to adeptly manage sequences of varying length while minimizing inference cost and grouped-query attention for swift inference (decrease inference time), details of which can be found in Ref. \cite{jiang2023mistral}.

\begin{table}
\begin{tabular}{ccccc}
\hline

&&Forward models&&\\
\hline
Prop/MAE&CFID&CGCNN&ALIGNN&AtomGPT\\
\hline
$E_{form}$ (eV/atom) & 0.142 & 0.063 &  \textbf{0.033} & 0.072 \\
OPT $E_g$ (eV) & 0.299 & 0.199 &0.142 & \textbf{0.139} \\
MBJ $E_g$ (eV)& 0.531 & 0.407 & \textbf{0.310} & 0.319 \\
$T_C$ (K)& 1.99 & 2.60 & 2.03 & \textbf{1.54} \\
\hline
&&Inverse models&&\\
\hline

&Database/RMSE&CDVAE& AtomGPT\\
\hline
&SuperConDB&0.24& \textbf{0.08}\\
&Carbon24&0.36&\textbf{0.32}\\
\hline

\end{tabular}
\caption{Performance measurement for forward and inverse design tasks and comparison with Classical Force-field Inspired Descriptors (CFID)\cite{choudhary2018machine}, Crystal Graph Convolutional Neural Network (CGCNN)\cite{xie2018crystal}, 
Atomistic Line Graph Neural Network (ALIGNN)\cite{choudhary2021atomistic} and Crystal Diffusion Variational Autoencoder (CDVAE)\cite{xie2021crystal} and AtomGPT models. Best performing model results are shown as bold texts. The forward property prediction (Prop.) models are shown using mean absolute error (MAE) metric between held out test set and property predictions, while the inverse models are compared using reconstruction loss in normalized root mean square error (RMSE) distance between held out test structures and those using the generative models.}

\end{table}

\begin{figure}[hbt!]
\centering
\includegraphics[width=\linewidth]{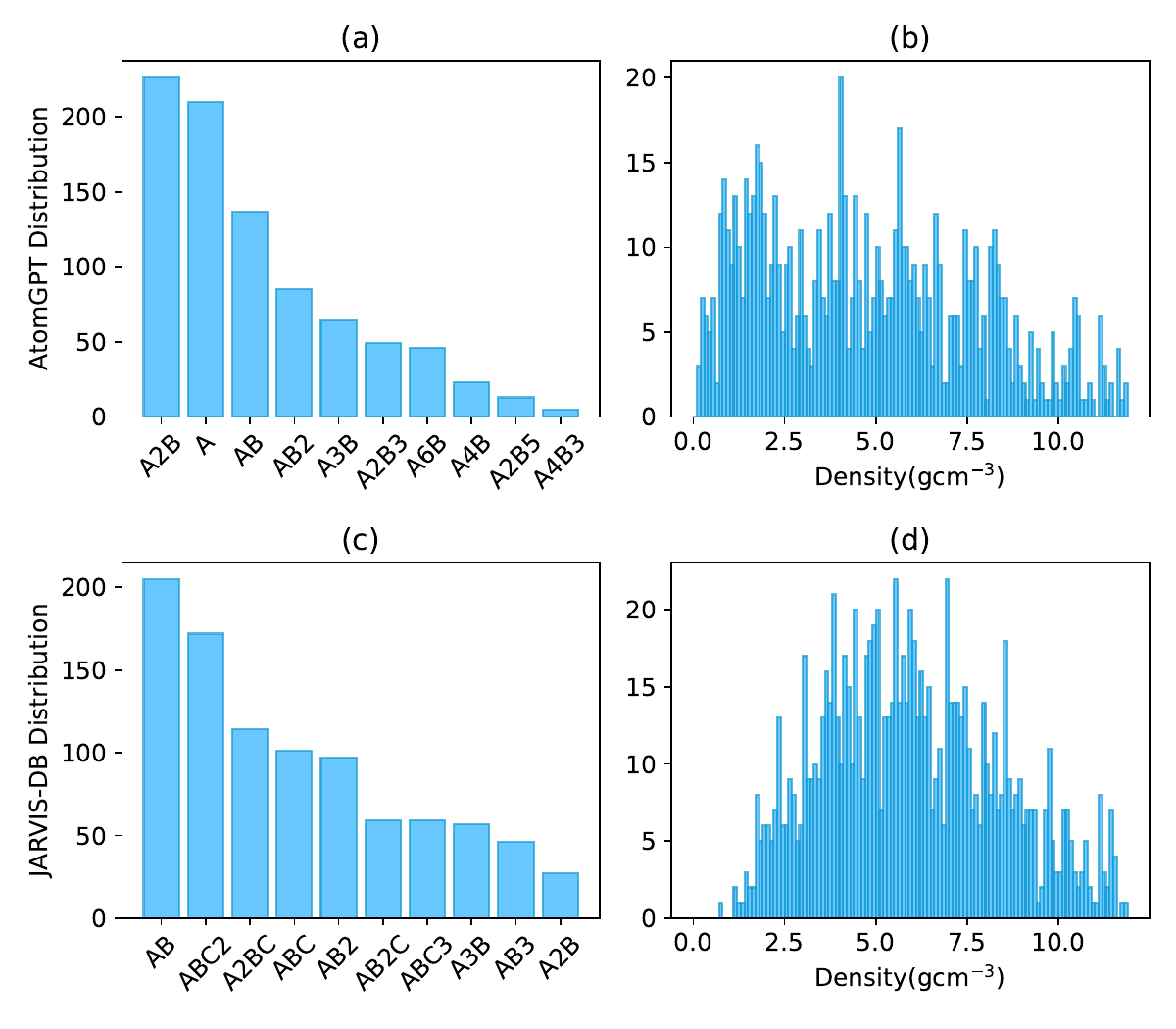}
\caption{ Chemical prototype and density of (a–b) the AtomGPT generated structures and (c–d) the 1058 structures used for training from the JARVIS-DFT database.}
\label{fig:zsl}
\end{figure}

Let us specifically focus on the superconductor inverse design task. This requires finetuning LLM models such as Mistral AI. Such finetuning requires transforming the instructions into a specialized protocol such as Alpaca \cite{taori2023stanford}. An example of converting MgB$_2$ into alpaca is shown in Fig. 1d. It consists of a python dictionary with Instruction, Input and Output keys. As directly finetuning such a model can be computationally expensive, Parameter-Efficient Fine-Tuning (PEFT) method using Hugging Face ecosystem (transformers, PEFT, Transformer Reinforcement Learning (TRL), Rotary Position Embedding (RoPE) \cite{su2024roformer} is employed in addition to patching a Mistral model with fast LoRA (Locally Reweighted Attention) \cite{hu2021lora} weights for reduced memory training. After obtaining the PEFT model, corresponding tokenizer, and alpaca dataset (with 80 \% training, 10 \% validation and 10 \% test splits for the superconductor dataset), supervised fine-tuning tasks are carried out with a batch size of 10,  AdamW 8 bit optimizer and cross-entropy loss function. This loss function measures the difference between the predicted probability distribution over the vocabulary and the true distribution (i.e., the one-hot encoded target words). After the model is trained, the model is evaluated on the test set with respect to reconstruction/test performance. To further clarify, after training the model on the train set, keeping the instruction and input keys in the test set, the trained model is employed to generate outputs. After parsing the outputs to create corresponding crystal structures,  the StructureMatcher algorithm \cite{ong2013python} is used to find the best match between two structures considering all invariances of materials. The root mean square squared error (RMSE) is averaged overall matched materials. Because the inter-atomic distances can vary significantly for different materials, the RMSE is normalized following the works in Ref. \cite{xie2021crystal}. Note that this is just one of the metrics for generative models for atomic structures and there can be numerous other types of metrics. It is important to calculate the $T_C$ using a high fidelity method such as DFT later that acts as a more reliable metric for the generation process. The performance of the model for the JARVIS-SuperconDB (1058) and Carbon24 (10153) datasets \cite{pickard2011ab} are shown in Table 1 and \textcolor{black}{Fig. 2q-2t}. We notice from Table 1 that AtomGPT has lower RMSE than CDVAE for both SuperconDB and Carbon24 databases suggesting that given a specific prompt, the model can generate structures very similar to the held out test set. \textcolor{black}{This can be further analyzed in Fig 2q-2t where the root mean square distance differences are lower for AtomGPT model than CDVAE models for both tasks. A higher number of bars at lower RMSE distance (d) values represent close agreement in structures.} \textcolor{black}{Moreover, the distributions for Fig. 2q and Fig. 2r are markedly different, while Fig. 2s and Fig. 2t display similar distributions. While there is no obvious explanation for such a behavior, a plausible explanation could be attributed to the inherent nature of the datasets used. The Carbon24 database is composed primarily of carbon-based structures generated via ab-initio random structure search (AIRSS) \cite{pickard2011ab}. Many of these entries are thermodynamically unstable and not synthesizable, which introduces higher variability in bond-distribution and low variability in chemical distribution and consequently higher RMSE in bond-distance values in the predictions. On the other hand, the SuperConDB contains a larger variety of diverse chemical compositions and stable structures. The stability and diverse nature of these structures could contribute to more consistent predictions and lower RMSE values, thus possibly explaining the observed difference in distributions. The similar distributions for carbon dataset (s-t) indicate that both modeling approaches are almost equally effective in predicting the structures within these datasets.} Further improvement could be possible with better text prompts, which are subject to future work. Note that only one property $T_C$ and specific type of instructions are used here, and LLMs allow easy addition of other properties and conditional statements as well.

\begin{figure}[hbt!]
\centering
\includegraphics[width=\linewidth]{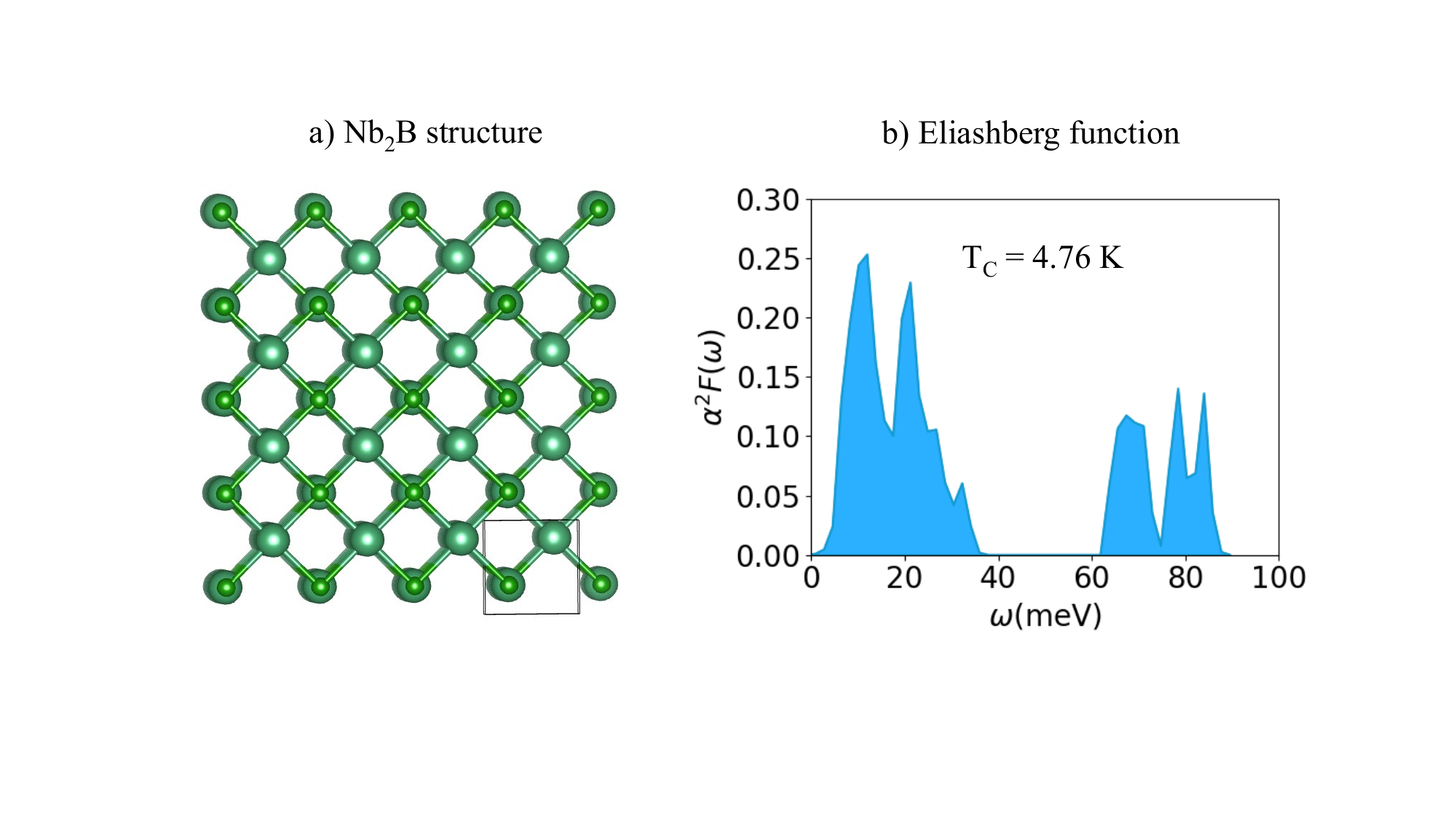}
\caption{Density functional theory calculation results for one of the identified structures. a) atomic structure visualization, b) Eliashberg function for Nb$_2$B structure with a predicted  $T_C$ of 4.76 K.}
\label{fig:nb2b}
\end{figure}

Next, materials are generated using such models for identifying potential candidate superconductors. Our focus in his work is on Borides (but are not limited by it) as they are one of the systems of interest for our future experimental synthesis. One of the first computational discoveries and experimentally synthesized superconductors is FeB$_4$ \cite{gou2013discovery} which further motivates such application. Systems with the formula X$_a$B$_b$, X$_a$Y$_b$B$_c$, X$_a$Y$_b$Z$_c$B$_d$, where X,Y,Z are periodic table elements other than Boron and a,b,c can be integers such as 1,2,3 are generated. After generating such 890 structures, the chemical prototypes and density distributions are shown in Fig. 3a and Fig. 3b. We observe that most of the generated structures belong to prototypes A2B, A, AB and AB$_2$. Their densities can attain values upto 15 gcm$^{-3}$. The prototypes generated are similar to those existing in the training dataset i.e., the JARVIS-DFT superconductor dataset as shown in Fig. 3c. The densities of the generated structures also closely resemble distributions of the training set with a peak around 5 gcm$^{-3}$ as shown in Fig. 3d. The best formation energy property model listed above is used to check if they have negative formation energy, bandgap models to make sure they have close to zero bandgaps and $T_C$ greater than 1 K leading to 30 candidate materials. Then, these candidates are subjected to ALIGNN-FF (developed in Ref. \cite{choudhary2023unified}) based pre-optimization. After the pre-optimization, these materials are subjected to JARVIS-DFT workflow using Vienna Ab initio Simulation Package \cite{kresse1996efficient,kresse1996efficiency} and OptB88vdW functional. Now, electron-phonon calculations are performed for these optimized structures. Note that this step is computationally expensive and can take days if not weeks for each material. One of the potential candidates is found to be Nb$_2$B with optimized lattice lengths of 3.19 {\textup{\AA}, 3.18 {\textup{\AA} and 4.52 {\textup{\AA} and lattice angles of 93.7\textdegree , 90\textdegree  and 90\textdegree crystallizing in the monoclinic spacegroup of P2/m (as shown in Fig. 4). It has a density of 7.1 gcm$^{-3}$ which is almost 3 times higher than Silicon (2.25 gcm$^{-3}$) and aluminum (2.66 gcm$^{-3}$). The formation energy and energy above the convex hull of this system are calculated as -0.13 eV/atom and 0.3 eV respectively. Finally, McMillan-Allen-Dynes formula based superconductor transition temperature calculations are performed using quantum espresso using workflow developed in Ref. \cite{choudhary2022designing} to calculate its $T_C$.

The workflow used to generate the training data is based on Ref. \cite{choudhary2022designing}, where  electron phonon calculations (EPC) were performed using non-spin polarized DFT-perturbation theory (PT) \cite{baroni1987green,gonze1995perturbation} (using the interpolated/Gaussian broadening method \cite{wierzbowska2005origins}) with the Quantum Espresso (QE) software package \cite{giannozzi2009quantum}, PBEsol functional \cite{perdew2008restoring}, and the GBRV \cite{garrity2014pseudopotentials} pseudopotentials. The EPC parameter is derived from spectral function ${\alpha}^2 F(\omega)$ which is calculated as follows:

\begin{equation} 
{\alpha}^2 F(\omega)=\frac{1}{2{\pi}N({\epsilon_F})}\sum_{qj}\frac{\gamma_{qj}}{\omega_{qj}}\delta(\omega-\omega_{qj})w(q)
\end{equation} 
where $\omega_{qj}$ is the mode frequency, $N({\epsilon_F})$ is the density of states (DOS) at the Fermi level ${\epsilon_F}$, $\delta$ is the Dirac-delta function, $w(q)$ is the weight of the $q$ point,  $\gamma_{qj}$ is the linewidth of a phonon mode $j$ at wave vector $q$ and is given by:

\begin{equation} 
\gamma_{qj}=2\pi \omega_{qj} \sum_{nm} \int \frac{d^3k}{\Omega_{BZ}}|g_{kn,k+qm}^j|^2 \delta (\epsilon_{kn}-\epsilon_F) \delta(\epsilon_{k+qm}-\epsilon_F)
\end{equation} 
Here, the integral is over the first Brillouin zone, $\epsilon_{kn}$  and $\epsilon_{k+qm}$ are the DFT eigenvalues with wavevector $k$ and $k+q$ within the $n$th and $m$th bands respectively, $g_{kn,k+qm}^j$ is the electron-phonon matrix element. $\gamma_{qj}$ is related to the mode EPC parameter $\lambda_{qj}$ by:

\begin{equation} 
\lambda_{qj}=\frac {\gamma_{qj}}{\pi hN(\epsilon_F)\omega_{qj}^2}
\end{equation} 
Now, the EPC parameter is given by:

\begin{equation} 
\lambda=2\int \frac{\alpha^2F(\omega)}{\omega}d\omega=\sum_{qj}\lambda_{qj}w(q)
\end{equation} 
with $w(q)$ as the weight of a $q$ point. The superconducting transition temperature, $T_c$ can then be approximated using McMillan-Allen-Dynes \cite{mcmillan1968transition} equation as follows:

\begin{equation}
T_c=\frac{\omega_{log}}{1.2}\exp\bigg[-\frac{1.04(1+\lambda)}{\lambda-\mu^*(1+0.62\lambda)}\bigg]\label{eq:mad}
\end{equation}
where
\begin{equation} 
\omega_{log}=\exp\bigg[\frac{\int d\omega \frac{\alpha^2F(\omega)}{\omega}\ln\omega}{\int d\omega \frac{\alpha^2F(\omega)}{\omega}}\bigg]
\end{equation} 
In Eq.~\ref{eq:mad}, the parameter $\mu^*$ is the effective Coulomb potential parameter, which we take as 0.1. It is important to note that the robustness of this workflow was heavily benchmarked against experimental data and higher levels of theory in ref. \cite{choudhary2022designing,2dsc,wines2023inverse,wines2023data}, which indicates that the training data for these deep learning models is of high quality, given the level of theory used to produce the data.

The $T_C$ for Nb$_2$B is predicted to be 4.76 K with no negative modes in the phonon density of states curve suggesting dynamical stability. Similar DFT calculations can be performed for other structures after populating more candidates, which can then be added to the JARVIS-DFT Superconductor database \cite{choudhary2022designing,2dsc,wines2023inverse,wines2023data}. Moreover, we can extend this workflow to include other candidate systems, such as nitrides and carbides. As the database of superconductors expands, both forward and inverse models have the potential for further improvement, which is necessary for enhanced predictions. After DFT validation, we can recommend these computationally discovered candidates for experimental synthesis. Superconductor design remains one of the most fascinating areas of research in physics and chemistry. Our work here serves as a proof-of-concept that demonstrates the potential of LLM models, such as GPTs, to accelerate superconductor materials design. Moreover, it can be easily extended to other classes of materials such as alloys, semiconductors, themoelectrics, dielectrics, piezoelectrics, solar cells and so on. Additionally, AtomGPT's application of AtomGPT for defect \cite{choudhary2023can} and interface \cite{D4DD00031E} materials design can also be interesting areas of research in the future.

\textcolor{black}{In conclusion, this study introduces AtomGPT, an atomistic generative pre-trained transformer model specifically designed for both forward and inverse materials design. The model is capable of predicting material properties with high accuracy and generating novel material structures with desired properties. By leveraging well-defined structural and chemical features from the JARVIS-Tools, AtomGPT ensures consistent and reliable input generation, enhancing the robustness of property predictions. The separation of forward and inverse models allows for efficient pre-screening and targeted structure generation, respectively. This dual approach facilitates the rapid discovery and optimization of new materials, paving the way for accelerated advancements in materials science. Note that, in this study, I present AtomGPT and large language models (LLMs) not as the ultimate solutions for materials science challenges, but as highly adaptable and tunable tools that can be tailored to a wide range of applications. The versatility of these models allows them to be customized for specific datasets and tasks, making them valuable complements to existing methodologies. By focusing on their adaptability and integration with other tools such as DFT and unified machine learning force-fields, I aim to highlight their potential to enhance and expand the toolkit available to materials scientists, facilitating innovation and discovery across the field. I have demonstrated the application of this workflow by computationally discovering a novel superconductor Nb$_2$B with $T_C$ of 4.76 K. These findings highlight the potential of large language models in transforming materials design processes, offering a promising direction for future research and applications in various domains of materials science.}

\section*{Acknowledgements}

K.C. thanks computational resources from the National Institute of Standards and Technology (NIST). This work was performed with funding from the CHIPS Metrology Program, part of CHIPS for America, National Institute of Standards and Technology, U.S. Department of Commerce. 
Certain commercial equipment, instruments, software, or materials are identified in this paper in order to specify the experimental procedure adequately. Such identifications are not intended to imply recommendation or endorsement by NIST, nor it is intended to imply that the materials or equipment identified are necessarily the best available for the purpose.

\bibliography{achemso-demo}

\end{document}